\documentclass{ws-mpla}
\usepackage[super]{cite}
\usepackage[utf8]{inputenc} 
\usepackage[T1]{fontenc}    
\usepackage{hyperref}       
\usepackage{url}            
\usepackage{booktabs}       
\usepackage{amsfonts}       
\usepackage{nicefrac}       
\usepackage{microtype}      
\usepackage{afterpage}
\usepackage{graphicx}
\usepackage{url}
\DeclareUnicodeCharacter{2212}{-}
\usepackage{color}
\usepackage{xcolor}
\usepackage{soul}

\begin{document}

\markboth{M. Telba and M. Binjonaid}
{Impact of LHC Higgs couplings measurements on bosonic decays of the neutral Higgs sector in the scNMSSM}

\catchline{}{}{}{}{}

\title{Impact of LHC Higgs couplings measurements on bosonic decays of the neutral Higgs sector in the scNMSSM}

\author{\footnotesize Marwa Telba\footnote{Both authors contributed equally to this work and share first authorship}}

\author{\footnotesize Maien Binjonaid$^{*}$\footnote{Correspondence: maien@ksu.edu.sa}}

\address{Department of Physics and Astronomy, King Saud University\\
Riyadh, Saudi Arabia}


\maketitle

\begin{abstract}
We analyze the Next-to-minimal supersymmetric standard model with Grand unification boundary conditions under current theoretical and experimental constraints. We compute the mass spectrum of the model and focus on the three lightest particles in the Higgs sector (two CP-even scalars, $h_1, h_2$, and one CP-odd, $a_1$). The reduced couplings of such particles, singlet-doublet components, their branching ratios to bosons, and reduced cross-section to photons and massive gauge bosons via gluon fusion are thoroughly and systematically scrutinized. Our analysis is focused on the parameter space where the singlet-doublet coupling $\lambda$ is as large as possible (keeping the perturbativity bound intact) and the ratio between the vacuum expectation values of the up-type and down-type Higgses ($\tan{\beta}$) is as small as possible, which is the region representing the most natural case of the NMSSM. We show the impact of recent constraints from the LHC on the SM-Higgs couplings to bosons and fermions on the parameter space of the model and the consequent implications on the Higgs sector. The results show that while the model is still able to account for current data, and provide an opportunity for discovery of extended Higgs sectors, recent LHC Higgs couplings constraints rule-out parts of the parameter space where $h_2$ (non-SM-like) and $a_1$ are non-singlet with masses below 400 GeV.

\keywords{NMSSM; Higgs Sector; LHC Constraints.}
\end{abstract}

\ccode{PACS Nos.: 14.80.Cp, 11.30.Pb, 12.60.−i}

\section{Introduction} \label{intro}

Since 2015, RUN II of the Large Hadron Collider (LHC) has been probing the frontiers of the Electroweak (EW) sector.
So far, around $139 \ fb^{-1}$ of data that is relevant to physics has been collected \cite{ATLAS-CONF-2019-021}. This data turns out to be consistent with the description of
the Standard Model (SM) of particle physics. Particularly, the unique scalar particle that was discovered
in 2012 \cite{Aad:2012tfa, Chatrchyan:2012xdj}, is consistent with the SM Higgs boson (See for e.g. \cite{Aad:2019mbh}), although its properties are still being considered \cite{Tanabashi:2018oca}. A crucial aspect of more precise measurement of the SM Higgs properties is to confront
experimental findings with the SM itself and Beyoned the SM (BSM) hypotheses which contain SM-like Higgs, such as supersymmetry (SUSY) \cite{Tanabashi:2018oca}.
The absence of any new physics at the LHC places stronger limits and erases portions of the
parameter spaces of models of new physics. Despite this fact, low-scale SUSY and models with a similar structure for the Higgs sector (e.g. 2HDM \cite{Branco:2011iw}) are still actively probed by the LHC. This is due to the fact that SUSY models still explain the
lightness of the Higgs mass (although with some degree of fine-tuning to compensate for its absence), and provide a candidate for particle dark matter, while passing all theoretical and experimental constraints \cite{vanBeekveld:2019tqp,Renga:2019mpg,Arbey:2018wjb}. 
There are many possible scenarios for SUSY to be part of the low-scale. The most considered one is the Minimal
Supersymmetry Standard Model (MSSM). However, it is well-known that the MSSM suffers from the $\mu-$problem \cite{Kim:1983dt}, and a large degree of fine-tuning (See for e.g. \cite{AbdusSalam:2019kei, vanBeekveld:2019tqp, Kobakhidze:2018vuy, Fowlie:2014xha, Kang:2012sy} and references therein).
One of the most natural class of SUSY models has been known for some time to be the Next-to-minimal supersymmetric standard mdoel (NMSSM) \cite{Ellwanger_2010,Maniatis:2009re}. Not only can the NMSSM solve the $\mu-$problem of the MSSM, it is also considerably less fine-tuned,
as has been demonstrated many times in the literature \cite{Barbieri:2006bg, Dermisek:2007yt,BasteroGil:2000bw,Hall:2011aa,PhysRevD.88.075003,Farina:2013fsa,Binjonaid:2014oga, Ellwanger:2014dfa, Athron:2017fxj}. 
Many aspects of the Higgs sector of the NMSSM has been considered in the literature \cite{Miller:2003ay, Wang:2020tap,Choi:2019yrv,Beskidt:2019mos, Beskidt:2017dil, Baum:2019uzg,King:2014xwa,King:2012tr,King:2012is, Cao:2018rix, Staub:2015aea, Ellwanger:2013ova, Ellwanger:2012ke, Potter:2015wsa, Cao:2012fz, Carmi:2012in}. It is well-known
that at tree-level, the NMSSM is able to predict a larger SM-Higgs mass than the MSSM, provided that $\tan{\beta}$ is small and $\lambda$ is large. However, $\lambda$ must not exceed $0.7$ if we care about perturbativity up the GUT scale \cite{Masip:1998jc, King:1995vk}.
With the LHC up and running, new Higgs-related limits and constraints have been placed on BSM models. For instance, the reduced couplings of the SM-Higgs boson to bosons and fermions provide a strong tool for constraining the parameter space of BSM scenarios. Combining recent Higgs
constraints to the large number of existing theoretical
and experimental constraints provides an interesting avenue
to study the allowed parameter space of the NMSSM, and analyze how exactly do recent constraints affect its parameter space.
In light of this, the aim of this paper is to study the implication of recent LHC constraints on the extended Higgs sector of the semi-constrained NMSSM (scNMSSM) at the low $\tan{\beta}$ and large $\lambda$ regime.
Particularly, we analyze the effects of such limits on
the mass range of the neutral Higgs bosons, their rare branching ratios to gauge bosons, their exotic decays to non-SM scalars, and the reduced cross-section to $\gamma \gamma/VV$ via gluon fusion.

The paper is organized as follows. In Section~\ref{sec:core1} we present a general introduction to the NMSSM, its input parameters, Higgs sector, and relevant quantities. Next, Section~\ref{sec:scans} is where we provide details about the methods used in this paper, and strategy of analyzing the parameter space of the model. In Section~\ref{sc:res}, the results of the paper are laid out. And finally, we discuss the results, and conclude our paper in Section~\ref{conc}.

\section{The NMSSM and relevant parameters}

\label{sec:core1}
The $\mu-$problem in the MSSM is tackled in the NMSSM by introducing a SM-singlet scalar,
along with its fermionic partner (the signlino). The singlet superfield couples with
the Higgs doublets. This coupling is encoded in the NMSSM superpotential \cite{Ellwanger_2010}, 
\begin{eqnarray}
W_{NMSSM} &=&h_{u}\widehat{Q}.\widehat{H}_{u}\widehat{U}_{R}^{c}+h_{d}%
\widehat{H}_{d}.\widehat{Q}\widehat{D}_{R}^{c}+h_{e}\widehat{H}_{d}.\widehat{%
L}\widehat{E}_{R}^{c}  \nonumber \\
&&+\lambda \widehat{S}\widehat{H}_{u}.\widehat{H}_{d}+\frac{1}{3}\kappa 
\widehat{S}^{3}.
\end{eqnarray}
Where $\widehat{Q}$ and $\widehat{L}$ denote the
left-handed doublet quark and lepton superfields while $\widehat{U}$, $%
\widehat{D}$ and $\widehat{E}$ represent the right-handed singlet up-type
quark, down-type quark and lepton superfields. The Yukawa couplings of the first and second generation fermions are ignored. Note that the
fourth term solves the MSSM $\mu $-problem by introducing an effective $\mu
- $term which is generated when the singlet superfield obtains a vacuum
expectation value (VEV) $\left\langle S\right\rangle =s$ \cite{Ellwanger_2010}.
The last term is introduced to break the Peccei-Quinn symmetry, to avoid
predicting an unobserved weak-scale massless Axion \cite{PhysRevLett.38.1440, PhysRevD.16.1791}.

The soft SUSY breaking term, which contains mass terms for all scalars ($%
m_{H_{u}}^{2}$, $m_{H_{d}}^{2}$, $m_{S}^{2}$, $m_{Q}^{2}$, $%
m_{U}^{2}$, $m_{D}^{2}$, $m_{L}%
^{2}$and $m_{E}^{2}$), gauginos ($M_{1},M_{2}$ and $%
M_{3})$ and the trilinear interaction terms ($A_{u},A_{d},A_{e
},A_{\lambda }$ and $A_{\kappa })$, can be expressed as \cite{Ellwanger_2010}, 
\begin{eqnarray}
-\mathcal{L}_{soft} &=&m_{H_{u}}^{2}\left\vert H_{u}\right\vert
^{2}+m_{H_{d}}^{2}\left\vert H_{d}\right\vert ^{2}+m_{S}^{2}\left\vert
S\right\vert ^{2}+m_{Q}^{2}\left\vert Q^{2}\right\vert   \nonumber \\
&&+m_{U}^{2}\left\vert U_{R}^{2}\right\vert +m_{D}^{2}\left\vert
D_{R}^{2}\right\vert +m_{L}^{2}\left\vert L^{2}\right\vert
+m_{E}^{2}\left\vert E_{R}^{2}\right\vert   \nonumber \\
&&+\frac{1}{2}\left[ 
\begin{array}{c}
M_{1}\lambda _{1}\lambda _{1}+M_{2}\sum_{i=1}^{3}\lambda _{2}^{i}\lambda
_{i2} \\ 
+M_{3}\sum_{a=1}^{8}\lambda _{3}^{a}\lambda _{a3}%
\end{array}%
\right]   \nonumber \\
&&+h_{u}A_{u}Q\cdot H_{u}U_{R}^{2}-h_{d}A_{d}Q\cdot
H_{d}D_{R}^{2}-h_{e}A_{e}L\cdot H_{d}E_{R}^{2}  \nonumber \\
&&+\lambda A_{\lambda }H_{u}\cdot H_{d}S+\frac{1}{3}\kappa A_{\kappa
}S^{3}+h.c.
\end{eqnarray}

One of the well-known aspects of the NMSSM is that the upper bound on the lightest
CP-even Higgs mass reads,
\begin{equation}
    m_h^2 \leq M_Z^2 \left( \cos^2{(2 \beta)} + \frac{\lambda^2}{g^2} \sin^2{(2 \beta)}  \right),  
\end{equation}
therefore, small $\tan{\beta} \leq 5$ and large $\lambda \leq 0.7$ is a viable option
to obtain the correct SM-Higgs mass without requiring loop contributions from the stop quarks,
as is the case in the MSSM.
At tree-level, the behavior of the NMSSM is specified by the following parameters, $$
m_0, \ m_{1/2}, \ A_{0}, \ A_{\lambda }, \ 
A_{\kappa }, \ \lambda ,\ \kappa , \ \tan{\beta},  \  \mu _{eff}. $$
The first three parameters represent the common scalar mass, the common
gaugino mass and the common trilinear coupling, respectively. $A_{\lambda }$
and \ $A_{\kappa }$ are the soft SUSY breaking trilinear NMSSM couplings.
These five parameters are GUT scale parameters. The coupling $\lambda $
represents the coupling between the two Higgs doublet and singlet
superfields while $\kappa $ corresponds to the singlet self-coupling. Here
tan$\beta $ is the ratio between the VEVs of the Higgs doublets, tan$%
\beta =\nu _{u}/\nu _{d}$, where $v^2 = v^2_u + v^2_d = (174 GeV)^2$, while $\mu _{eff}$ is defined via the VEV of the
Singlet field $s$ to be $\mu _{eff}=\lambda s$. The last four parameters are given at the low scale.

The extended Higgs sector in the NMSSM is much richer than the MSSM in that it consists
of seven Higgs particles. The mass matrices contain a new singlet component, therefore
the properties of the NMSSM Higgs sector can be very different from that of the MSSM.
Moreover, the neutral scalar fields in the Higgs sector couple at tree-level to massive bosons and fermions. They also have loop-induced
couplings to photons and gluons. Such couplings determine the possible decays and production channels.
Relative to the SM, one denotes the reduced couplings of a neutral Higgs to gauge bosons, up-type and down-type fermions by: $C_{V}$, $C_{U}$, $C_{D}$, respectively. Additionally, reduced couplings to gluons $C_{g}$, and photons $C_{\gamma}$ are to be considered since they are affected by contributions from new physics. As mentioned, these reduced couplings are defined by dividing a given coupling as predicted in the NMSSM
by the corresponding SM prediction. A thorough description of such couplings is given in Refs. \cite{Belanger:2012gc, Ellwanger:2012ke, Belanger:2013xza}.

In this study, and to limit our analysis, we focus on the important decays of the type $\phi \rightarrow$ bosons, where $\phi$ denotes $h_1, h_2$ (CP-even) or $a_1$ (CP-odd). Furthermore, the production of neutral Higgs bosons takes place via several channels, the most important one being the gluon fusion channel (ggF) \cite{King:2012tr}. A useful quantity to understand how a signal from a given Higgs particle would differ from the SM Higgs is the so-called reduced cross-sections, defined as \footnote{NMSSMTools, see the next Section, which is the package we utilize uses this definition.},  
\begin{equation}
R_{j}^{i}=C_{i}^{2}\frac{\left ( BR_{j} \right )_{NMSSM}}{\left ( BR_{j} \right )_{SM}}.
\end{equation}
It is calculated through the reduced couplings $C_i$, the index $i$ denotes the couplings associated with a given production channel.
In our case, we will be focusing on gluon fusion, since it is known to give the largest contribution to the
production of the Higgs as mentioned above. The index $j$ denotes the decay channel.

\section{Scan Strategy} \label{sec:scans}
To analyze the parameter space of the NMSSM, we utilize the state-of-the-art version of $\mathtt{NMSSMTools \ v.5.5.2}$  \cite{NMSSMTools,Ellwanger:2004xm,Ellwanger:2005dv,Das:2011dg,Muhlleitner:2003vg} and modify it to our purpose. 
We have considered a constrained type of GUT-scale boundary conditions on the gaugino masses, the scalar masses.
In our scans, we consider the case where $A_{\lambda}$ and $A_{\kappa}$ are not equal to $A_0$. Given that the two trilinears are specific to the NMSSM, it is interesting to analyze their effect on the parameter space of the model.
NMSSMTools allows one to specify some parameters at the GUT-scale, while some parameters are specified at the low-scale. Particularly, $m_0, m_{1/2}, A_0, A_{\lambda}, A_{\kappa}$ are specified at the GUT-scale, whereas $\tan{\beta}$ is specified at $M_Z$, and $\lambda, \kappa$ are specified at the SUSY scale, defined by the masses of the first generation of squarks.  It is worth mentioning here that for the previous set to be allowed as inputs in NMSSMTools, both $m_{H_i}$ ($i=u,d$) are computed at the GUT scale. Therefore, technically, this is called the semi-constrained NMSSM, since the Higgs mass parameters are not equal to $m_0$ at the GUT scale.

The range of parameters our scans are:
\begin{eqnarray*}
& m_0(\text{GUT})=[500-4000] GeV, 
& m_{1/2}(\text{GUT})=[500-4000] GeV,  \\
& (A_0,A_{\lambda},A_{\kappa})(\text{GUT})=[-3000-3000] GeV, 
& \tan{\beta}(M_Z)= [1-10]  \\
& (\lambda,\kappa)(\text{SUSY}) =[0.4,0.3-0.7], 
& \mu_{eff}(\text{SUSY})=[100-1500] GeV.
\end{eqnarray*}
Within these ranges, we randomly scanned up to $10^8$ points to have a good representation of the semi-constrained NMSSM.

In effect, a total of $75$ types of constraints are implemented in NMSSMTools. In our analysis, we
treat Higgs related constraints separately from non-Higgs related ones, as will be explained shortly. Non-Higgs constraints include theoretical ones (non-tachyonic masses, successful EWSB, the existence of global minimum),
and phenomenological ones (flavor physics, LHC constraints
on sparticles, satisfying the upper limit on dark matter relic density) \footnote{For a comprehensive list of all implemented limits, we refer the reader to the official webpage of NMSSMTools.}. We let NMSSMTools eliminate any point that suffers from a problem with regards to those limits.
On the other hand, Higgs-related constraints are pre-LHC and LHC. Pre-LHC limits include, for e.g., LEP, and ALEPH results, while LHC constraints include the very important limits on the reduced couplings. The limits implemented in NMSSMTools are a combination of recent ATLAS and CMS results (See references in Sec.~\ref{intro}).

A crucial type of constraints on the Higgs sector come from the process $aa \rightarrow XY$, where $X$ and
$Y$ could be two pairs of leptons (e.g. 4$\mu \bar{\mu}$), or a pair of leptons and a pair bottom quarks,
or two pairs of bottom quarks. LHC constraints on such decays are implemented in NMSSMTools.
All issues associated with the process $aa \rightarrow XY$ are grouped into one problem denoted by $h \rightarrow aa \rightarrow 4l/2l+2b$.

We have applied all related constraints while allowing the passage of points that have problems with regards to Higgs couplings. These points are collected and binned into files according to the type of problem associated with each point. Points that have no issues are collected
separately as "good points". Our goal in doing so is to systematically
explore the effects of Higgs couplings' constraints on the parameter space and
predictions of the model. 

\section{Numerical results} \label{sc:res}
Here we present our results after imposing recent LHC constraints.
We focus on the lightest three neutral Higgs particles in this sector. Namely, $h_1, h_2$ (CP-even)
and $a_1$ (CP-odd). As the spectrum is calculated, we show
the effects of LHC SM-Higgs couplings' constraints on the parameter space. The ranges of input parameters after imposing all constraints is quite similar to that specified in Section~\ref{sec:scans}, except that the maximum value of $\kappa$ is close to 0.6. In the Appendix, we provide the details of a representative point in the parameter space passing all constraints and having a value of $m_{a_1} \approx 65$ GeV.


\subsection{Reduced couplings } 
\begin{figure*}
\centering
      \includegraphics[width=0.85\textwidth]{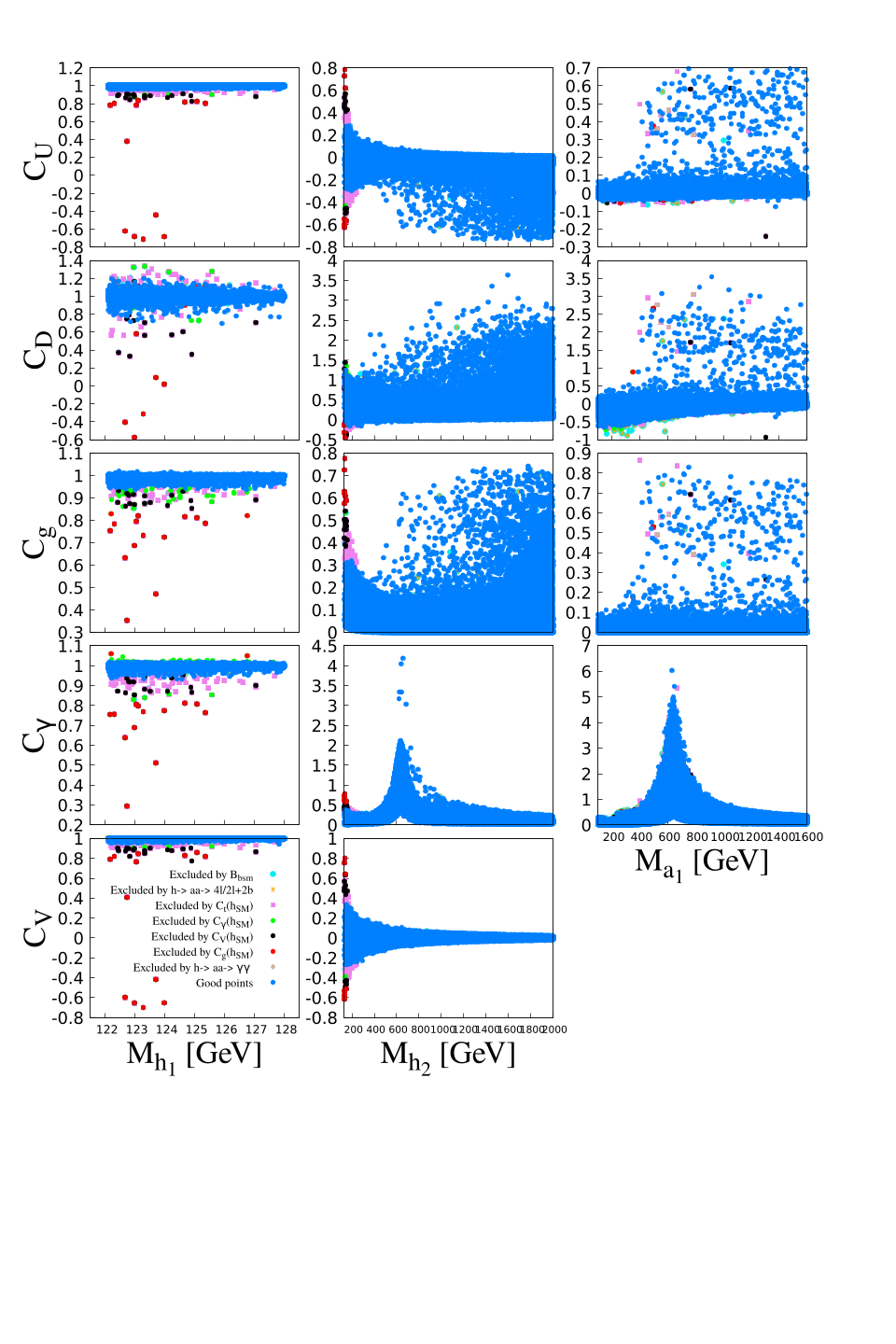}
    \caption{Reduced couplings of $h_{1}$ (left panels), $h_{2}$ (middle panels) and $a_{1}$ (right panels) to up-type and down-type fermions, glouns, photons and gauge bosons, respectively for $A_{0}\neq A_{\lambda }\neq A_{\kappa }$.}
    \label{Rcouplingh1h2a1case2}  
\end{figure*}

Fig.~\ref{Rcouplingh1h2a1case2} presents the results of the reduced couplings for $h_1$, $h_2$, and $a_1$. The results show that, in the scanned parameter space, the lightest CP-even Higgs, $h_{1}$, turns out to be the SM-like Higgs. Therefore, it is expected that the reduced couplings of $h_1$ are close or equal to unity. Points that pass or violate LHC constraints on these reduced couplings are shown in colors. The points in the violet square, green circle, black circle, red circle, and cyan circle are ruled-out due to being $2\sigma$ away from the LHC measured values for$C_{t}(h_{SM})$, $C_{\gamma}(h_{SM})$, $C_{V}(h_{SM})$, $C_{g}(h_{SM})$ and $B_{BSM}(h_{SM})$, respectively. On the other hand, points in orange star, and beige circle are ruled-out due to the constraints from $h\rightarrow aa\rightarrow 4l/2l+2b$ and $h\rightarrow aa\rightarrow \gamma \gamma $. Lastly, the surviving points passing all constraints are presented in the blue point circle.

We observe in the first row, second column of Fig.~\ref{Rcouplingh1h2a1case2} that, the reduced coupling of $h_{2}$ to up-type quarks, $C_U$, can vary between -0.3 and 0.3 for mass values below 400 GeV. The impact of LHC SM-Higgs couplings' constraints can be seen in this region, while the positive values decrease to below 0.1 for the rest of the points. On the other hand, the negative values reach up to -0.7 for a mass value above 1 TeV. The ruled-out points have $C_t(h_{SM}),$ $C_{V}(h_{SM})$, $C_{\gamma}(h_{SM})$, and $C_{g}(h_{SM})$ outside the experimentally allowed values as can be seen in the Figure. Next, the reduced coupling of $h_2$ to down-type quarks, $C_D$, takes values between 0 and just below 4. In most of the allowed parameter space, it takes values between 0 and 2. The ruled-out points are shown in the second row, second column in Fig.~\ref{Rcouplingh1h2a1case2}. The impact of SM-Higgs couplings' constraints are concentrated in the region where $m_{h_2} < 300$ GeV. As for the reduced coupling of $h_2$ to gluons, $C_g$, we see that it is 0.3 maximum for mass values near 200 GeV, while for larger masses it can reach 0.7. The ruled-out points violate the constraints on: $C_t(h_{SM}),$ $C_{V}(h_{SM})$, $C_{\gamma}(h_{SM})$, and $C_{g}(h_{SM})$ which is clearly shown in the third row, second column of Fig.~\ref{Rcouplingh1h2a1case2}. Furthermore, the reduced coupling of $h_2$ to photons, $C_{\gamma}$, is peaked at 4 for a mass value of 600 GeV, and drops sharply for other values of mass to 0.5 and below. A small region, shown in Fig.~\ref{Rcouplingh1h2a1case2} (forth row, second column), is ruled-out due to being associated with $C_g(h_{SM}),$ and $C_V(h_{SM})$ that are more than $2\sigma$ away from the measure values at the LHC. Lastly for $h_2$, the last row, second column of Fig.~\ref{Rcouplingh1h2a1case2}, shows that the coupling to vector bosons $C_V$ is almost always between -0.1 and 0.1 for mass values above 400 GeV, while it can be between -0.3 and 0.3 for smaller values of mass. The ruled-out points come with values of $C_t(h_{SM}),$ $C_{V}(h_{SM})$, and $C_g(h_{SM})$ that are outside what is experimentally allowed by more than $2\sigma$.

Advancing to the last considered particle, $a_{1}$, where the results of its reduced couplings are presented in the third column of Fig.~\ref{Rcouplingh1h2a1case2}. In most of the allowed parameter space, the reduced couplings $C_U, C_D$ and $C_g$ range from 0 and 0.05, -0.5 to 0.5, and 0 to 0.1, respectively. However, the value can vary up to 0.7, 3.5, and 0.8, respectively for $m_{a_1} > 400$ GeV. Finally, $C_{\gamma}$ peaks at 6 for a $m_{a_1}$ slightly larger than 600 GeV, while it drops sharply below 1 for $m_{a_1} < 500$ GeV or $m_{a_1} > 900$ GeV. These large values and those in the case of $h_2$ occur in the region where the effective coupling of the SM Higgs to photons in the denominator is calculated for a SM Higgs mass of around 600 GeV, and there the value of the coupling drops sharply due to cancellations between loop contributions (for more details see Ref.~\cite{Djouadi:2005gi}). The ruled-out points are randomly scattered in the $a_1$ data. However, a clear region that has been impacted by LHC SM-Higgs constraints is seen the second row, third column of Fig.~\ref{Rcouplingh1h2a1case2}, where values of $C_D$ that are below -0.5 ($m_{a_1} < 600$ GeV) are associated with points that violate the contrarians, especially on $B_{BSM}(h_{SM})$, $C_{\gamma}(h_{SM})$.

\subsection{Doublet and singlet components }

\begin{figure*}
\centering
      \includegraphics[width=1\textwidth]{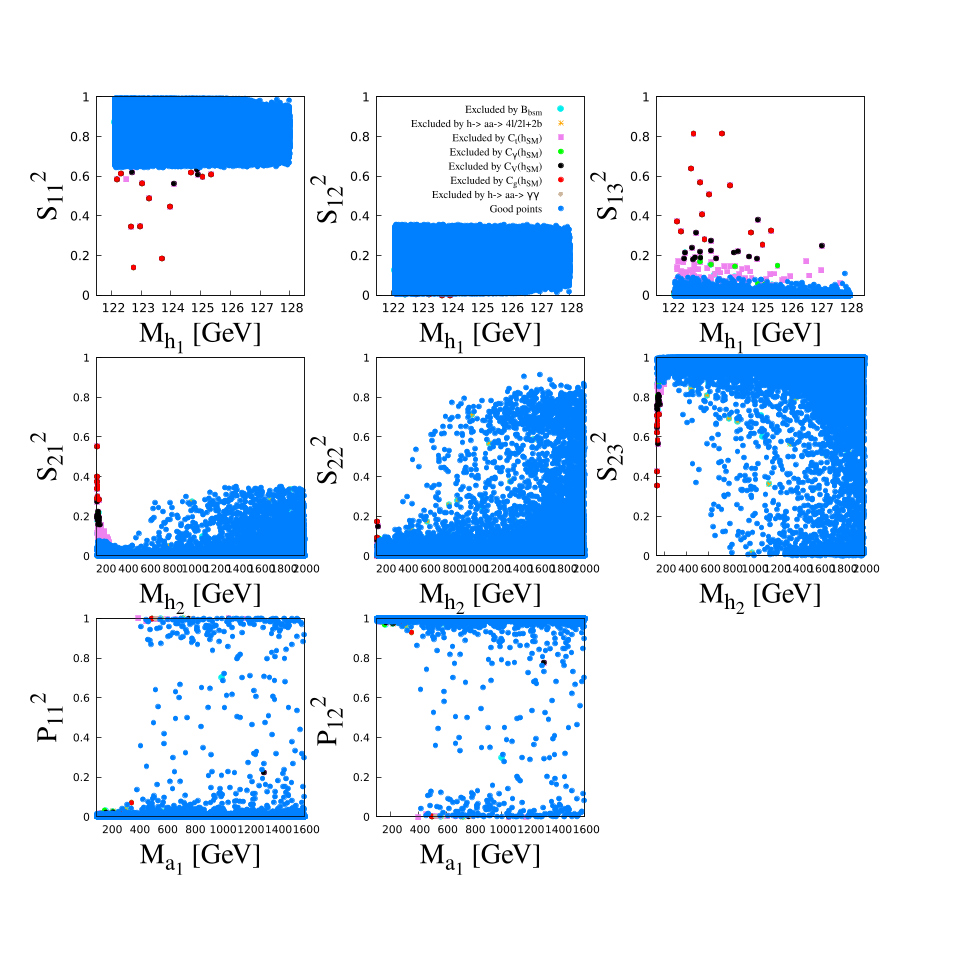}
    \caption{ Singlet and Doublet components of the lightest CP-even Higgs boson $h_{1}$ (upper panels), the next to lightest CP-even Higgs boson $h_{2}$ (middle panels) and the lightest CP-odd Higgs boson $a_{1}$ (lower panels) with $\mathbf{A}_{0}\neq \mathbf{A}_{\protect\lambda
}\mathbf{\neq A}_{\protect\kappa }$.}
    \label{D&SCompH1H2A1case2}  
\end{figure*}

Fig.~\ref{D&SCompH1H2A1case2} shows the doublet and singlet components of the three lightest Higgs bosons in the NMSSM framework contributing the its mass. For the CP-even Higgs states, $h_{1}$ and $h_{2}$, $S_{i1}$ and $S_{i2}$ represent the doublet components of the weak eigenstates, $H_{u}$ and $H_{d}$, respectively, whereas $S_{i3}$ represents the singlet component. For the CP-odd Higgs state, $a_{1}$,  $P_{11}$ and $P_{12}$ represent the doublet and singlet-component of this boson. Starting with the lightest CP-even Higgs state, the leftmost side of this Figure shows that, for the allowed points, the first doublet component $S^2_{11}$ varies from $0.6$ up to $0.99$ which gives a large contribution. On the middle plot, the component $S_{12}^2$ ranges from $0$ to $0.36$, while the singlet component $S_{13}^2$ takes a maximum value just below $0.1$, signaling that in all of the allowed parameter space the singlet contribuition to $h_1$ is very small. The ruled-out points are $2 \sigma$ away from the LHC constraints on $C _{t}(h_{SM})$, $C_{V}(h_{SM})$, $C_{\gamma}(h_{SM})$, and $C_{g}(h_{SM})$.
As seen in the middle panel of the Figure, $h_{2}$ can be mostly singlet for all ranges of allowed mass. In fact, the rightmost side indicates that where the mass is below $400$ GeV, this particle is always mostly singlet. Above that mass, the singlet component can reach zero. On the other hand, the doublet components of $h_2$, $S_{21}^2$ and $S_{22}^2$, reach maximum values of $0.4$ and $0.9$, respectively, where $m_{h_2} > 400$ GeV. The figure also shows that LHC Higgs limits rule-out regions where the mass of $h_2$ is smaller than 200 GeV with a singlet component below 0.9. Specifically, this region (shown as violet, black and red points) is associated with points that violate
the LHC constraint on $C _{t}(h_{SM})$, $C _{V}(h_{SM})$, and $C _{g}(h_{SM})$ (by more than $2\sigma$). 

Finally, as shown in the lower part of this Figure, the doublet component $P_{11}^2$ of $a_{1}$ is always close to zero for ranges of mass below 400 GeV, whereas it can reach values close to 1 for masses above that. On the other hand, the singlet component can be close to unity for all ranges of mass, and $a_1$ is only singlet for masses below 400 GeV. 
The impact of LHC Higgs constraints are visible
in the region where $m_{a_1} < 300$ GeV. Regions where $m_{a_1} \leq 61$ GeV are ruled-out by the combination of constraints on
$C _{V}(h_{SM})$ and $C _{\gamma}(h_{SM})$. 

\subsection{Branching ratios and reduced cross-sections} \label{Brc2}

\subsubsection*{Lightest CP-even Higgs ($h_1$)}

\begin{figure*}
\centering
      \includegraphics[width=1\textwidth]{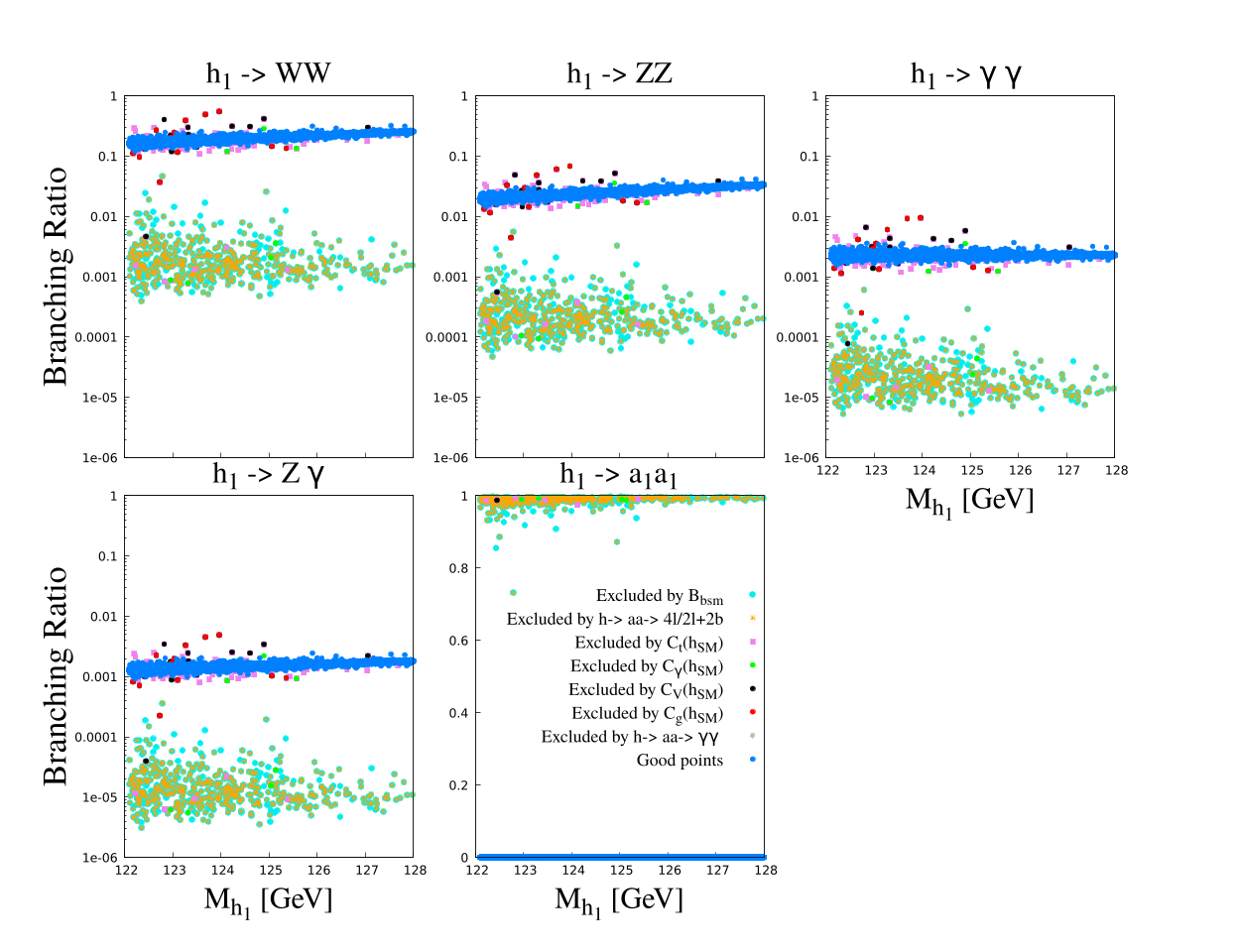}
    \caption{The branching ratios of $h_{1}\rightarrow WW$ , $h_{1}\rightarrow ZZ,$ $h_{1}\rightarrow \gamma \gamma ,$ $h_{1}\rightarrow Z\gamma $ and $%
h_{1}\rightarrow a_{1}a_{1} $ plotted aginest the lightest
CP-even Higgs mass $m_{h_{1}}$ for $A_{0}\neq A_{\lambda }\neq A_{\kappa }$.}
    \label{BRH1CASE2}  
\end{figure*}

As previously stated, the scanned parameter space is associated with $h_1$ being the SM-like Higgs, with mass range: $122$ GeV $ < m_{h_1} < 128$ GeV to account for theoretical uncertainties.
As expected, in the allowed regions shown in blue color, the branching ratios of $h_1$ are similar to the SM as presented in Fig.~\ref{BRH1CASE2}. Namely, $ h_1 \rightarrow WW$ slightly varies from 0.1 for the lower end of the predicted mass to 0.3 for the higher end. Near the allowed band (in blue) it can be seen that some points violate the implemented constraints.
More notably, regions where branching ratio is already well below the accepted value in the SM are also associated with violations of the limits of $B_{BSM}(h_{SM})$ and $h\rightarrow aa\rightarrow 4l/2l+2b$. Next, we consider $ h_1 \rightarrow ZZ$, where the allowed band (in blue) varies from 0.01 for the lower end of mass to 0.04 at the higher end, which is a much smaller separation than that for the decay to W bosons case. As for $ h_1 \rightarrow \gamma \gamma$ and $ h_1 \rightarrow \gamma Z$, the allowed band showing the predicted branching ratio levels at 0.005. Finally, $ h_1 \rightarrow a_1 a_1$ is not a successful decay channel in the presented parameter space. As can be seen in the Figure, values of the branching ratio near 1 are associated with ruled-out points due to $B_{BSM}(h_{SM})$ as well as $ h_1 \rightarrow aa \rightarrow 4l/2l+2b$. Indeed these two limits have the most impact on the parameter space.

The top row of Fig. \ref{REDh1h2a1case2} indicates that the
reduced cross-section of the lightest CP-even Higgs lays around unity, which is expected since it represents the SM-like Higgs in our scanned parameter space. It is worthwhile to mention that some of the excluded points (where $R>1.5$) violate the limits on $C _{t}(h_{SM})$, $C _{V}(h_{SM})$ and $C _{g}(h_{SM})$. 

\subsubsection*{Second lightest CP-even Higgs ($h_2$)}

\begin{figure*}
\centering
      \includegraphics[width=0.8\textwidth]{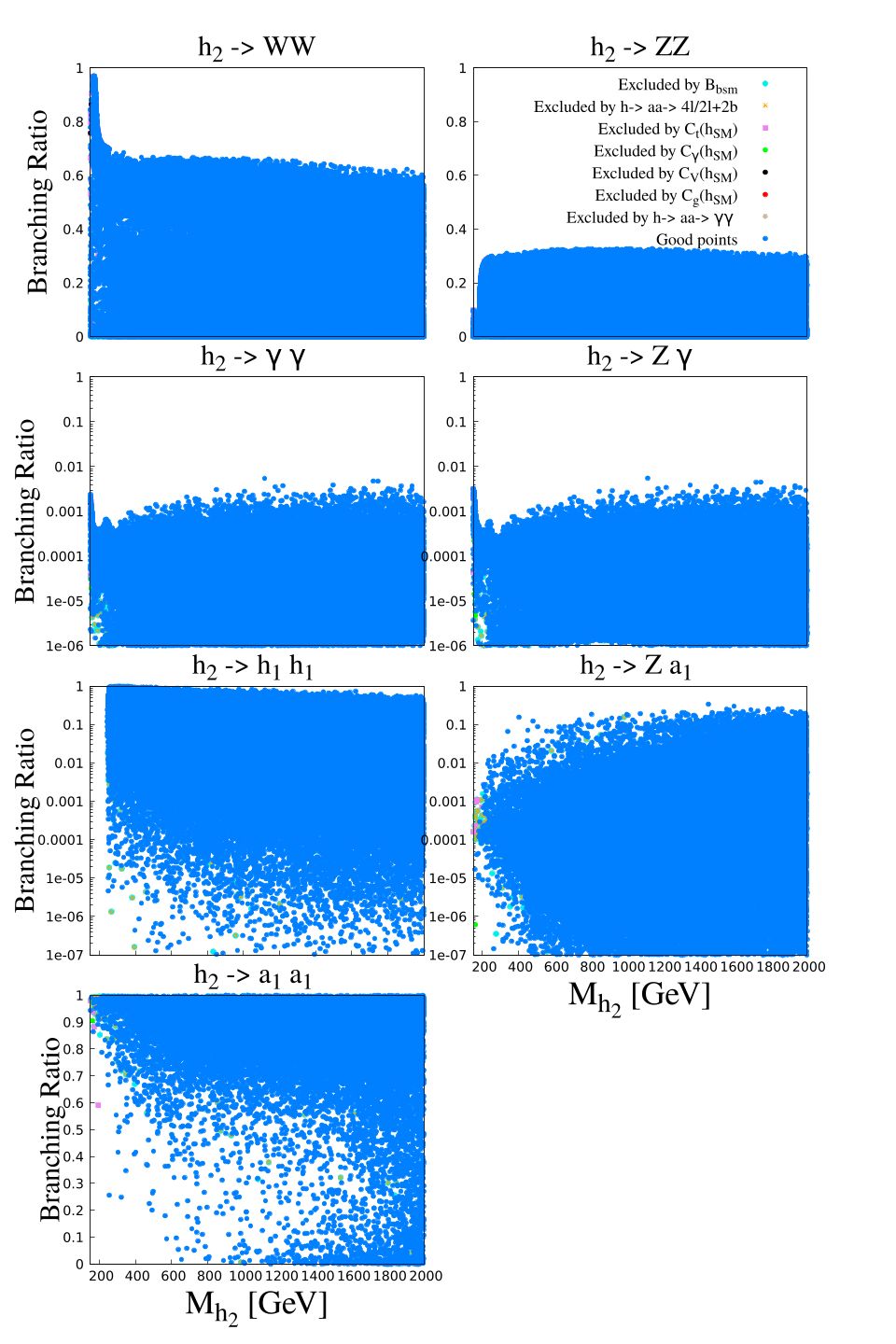}
    \caption{The branching ratios of $h_{2}\rightarrow WW,$ $h_{2}\rightarrow ZZ,$ $h_{2}\rightarrow \gamma \gamma ,$ $h_{2}\rightarrow Z\gamma $, $h_{2}\rightarrow
h_{1}h_{1}$, $h_{2}\rightarrow Za_{1}$  and $h_{2}\rightarrow a_{1}a_{1}$ plotted aginest the next to
lightest CP-even Higgs mass $m_{h_{2}}$ for $A_{0}\neq A_{\lambda }\neq A_{\kappa }$.}
    \label{BRH2CASE2}  
\end{figure*}

The branching ratios of $h_2$ are shown in Fig.~\ref{BRH2CASE2}.
Beginning with $h_2 \rightarrow WW$, it is clear that $h_2$ decays can be dominant to $WW$ for values of mass below 200 GeV. In that region, $h_2$ is mostly singlet as presented in the previous Section. As the mass becomes larger than 200 GeV, the maximum value of the branching ratio plateaus close to 0.65. 
Next, the plot for $h_2 \rightarrow ZZ$ indicates that the maximum value for the branching ratio is about 0.3 for all ranges of mass except for a small region where $m_{h_2} < 200$ GeV. In that area, $Br(h_{2} \rightarrow ZZ) < 0.3$ always.  
Moreover, the decays $h_2 \rightarrow \gamma \gamma, Z \gamma$ have similar properties. The peak value of branching ratio is around 0.0011 for the low mass range. However, the lowest value of the branching ratio in that region is just below $1 \times 10^{-5}$. Below that, problems associated with the $h_{SM}\gamma\gamma$ coupling arise.
An interesting decay is $h_2 \rightarrow h_1 h_1$, where for $m_{h_2} > 256$ GeV, this particle can predominantly decay into SM-like Higgs.
As for $h_2 \rightarrow Z a_1$, we note that this channel reaches a maximum value of $Br(h_{2} \rightarrow Z a_{1}) \sim 0.1$ for $m_{h_2} > 300$ GeV. In associated plot, we also note that the effects of SM-Higgs coupling constraints are visible in the region where $m_{h_2} < 250$ GeV, and $ 0.0001 < Br < 0.0011$. There, the limits on $h \rightarrow aa \rightarrow 4l/2l+2b$, $C _{t}$, and $B_{BSM}(h_{SM})$ are responsible for ruling out any predictions.
For the last decay channel, it is clear that $h_2$ can decay mainly to $a_1 a_1$ for all ranges of its mass. As $m_{h_2}$ increases, the chance that this channels becomes less significant (or even negligible) increases. The impact of the constraints are more visible in the region where the mass is below 200 GeV and $Br < 1$. This region suffers from the same aforementioned problems.

The reduced cross-section $R(ggF \rightarrow h_{2} \rightarrow \gamma \gamma)$ is shown in the middle part of Fig.~\ref{REDh1h2a1case2}, where the peak value ($95$) is reached at $m_{h_2} = 656$ GeV. Points where $m_{h_2} < 600$ GeV are associated with very small reduced cross-sections. For very small $m_{h_2}$, the effect of the constraint on the coupling of the SM-like Higgs to gluons is present. 
Finally, $R(ggF \rightarrow h_{2} \rightarrow WW/ZZ)$ clearly shows that the considered coupling constraints rule-out values of reduced cross-section above 0.1 for this channel. 
All in all, we observe that SM Higgs couplings constraints affected regions where $m_{h_{2}} < 300$ GeV in the considered parameter space. 

\subsubsection*{Lightest CP-odd Higgs ($a_1$)}

\begin{figure*}
\centering 
      \includegraphics[width=1\textwidth]{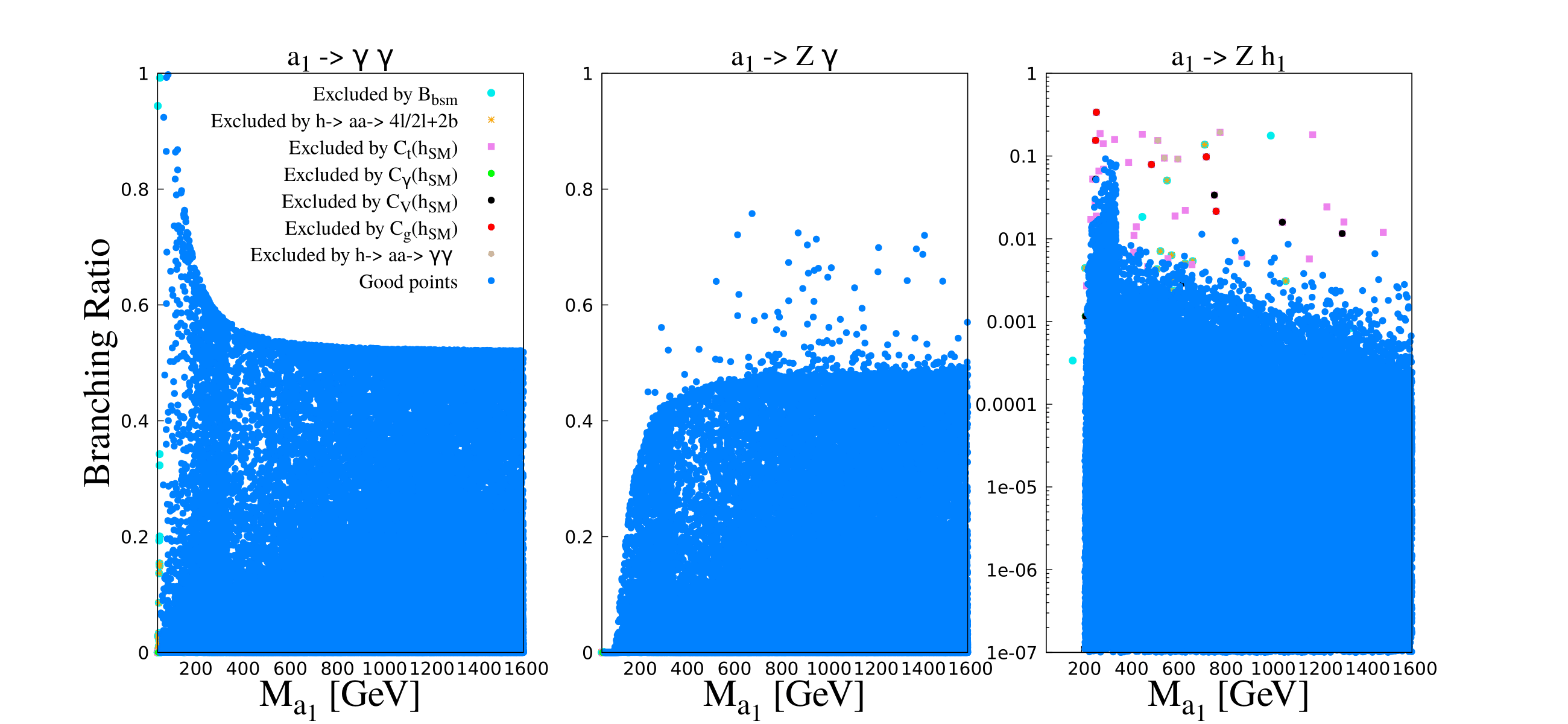}
    \caption{The branching ratios of $a_{1}$ $\rightarrow \gamma
\gamma ,$ $a_{1}\rightarrow Z\gamma ,$ and $a_{1}\rightarrow Zh_{1}$ plotted aginest the lightest
CP-odd Higgs mass $m_{a_{1}}$ for $A_{0}\neq A_{\lambda }\neq A_{\kappa }$.}
    \label{BRA1CASE2}  
\end{figure*}

In the allowed parameter space, the mass of $a_1$ ranges from $61$ GeV to $3.4$ TeV. 
Fig.~\ref{BRA1CASE2} lays out the results for the branching ratios of $a_1$ to bosons. Firstly, the maximum value of $Br(a_1 \rightarrow \gamma \gamma)$ varies between 0.5 and 0.9, where the latter is reached when $m_{a_1} \sim 150$ GeV.
Secondly, the branching ratio to $Z\gamma$ starts close to zero for $m_{a_1} \rightarrow 0$, then it sharply increases to about 0.4 for $m_{a_1} \sim 300 $ GeV, and it eventually levels at about 0.5 for larger masses. However, there are successful points where $Br(a_1 \rightarrow Z \gamma)$ reaches a value just below 0.8, as seen in the plots. Thirdly, the branching ratio of the channel $a_1 \rightarrow Z h_1$ is only non-zero for $m_{a_1} > 200$ GeV. It peaks at a value of 0.1 for $m_{a_1} \sim 350$ GeV, then drops sharply to 0.01 until $m_{a_1} \sim 400$ GeV. After that, the value of the branching ratio drops linearly to 0.001 as $m_{a_1} \rightarrow 1.6$ TeV. The rightmost plot of this Figure shows that the ruled-out points violate all of the considered SM-Higgs couplings constraints. It is worth mentioning that the large values of the branching ratio appearing in the Figure take place when both $a_1$ and $m_{h_2}$ are mainly singlets (i.e. singlet components close to unity for both). In such a case, the branching ratio can be enhanced as the results show, and this has been known in the literature (e.g. the recent analysis in Ref.~\cite{Almarashi:2018fgw}).

\begin{figure*}
\centering
      \includegraphics[width=1\textwidth]{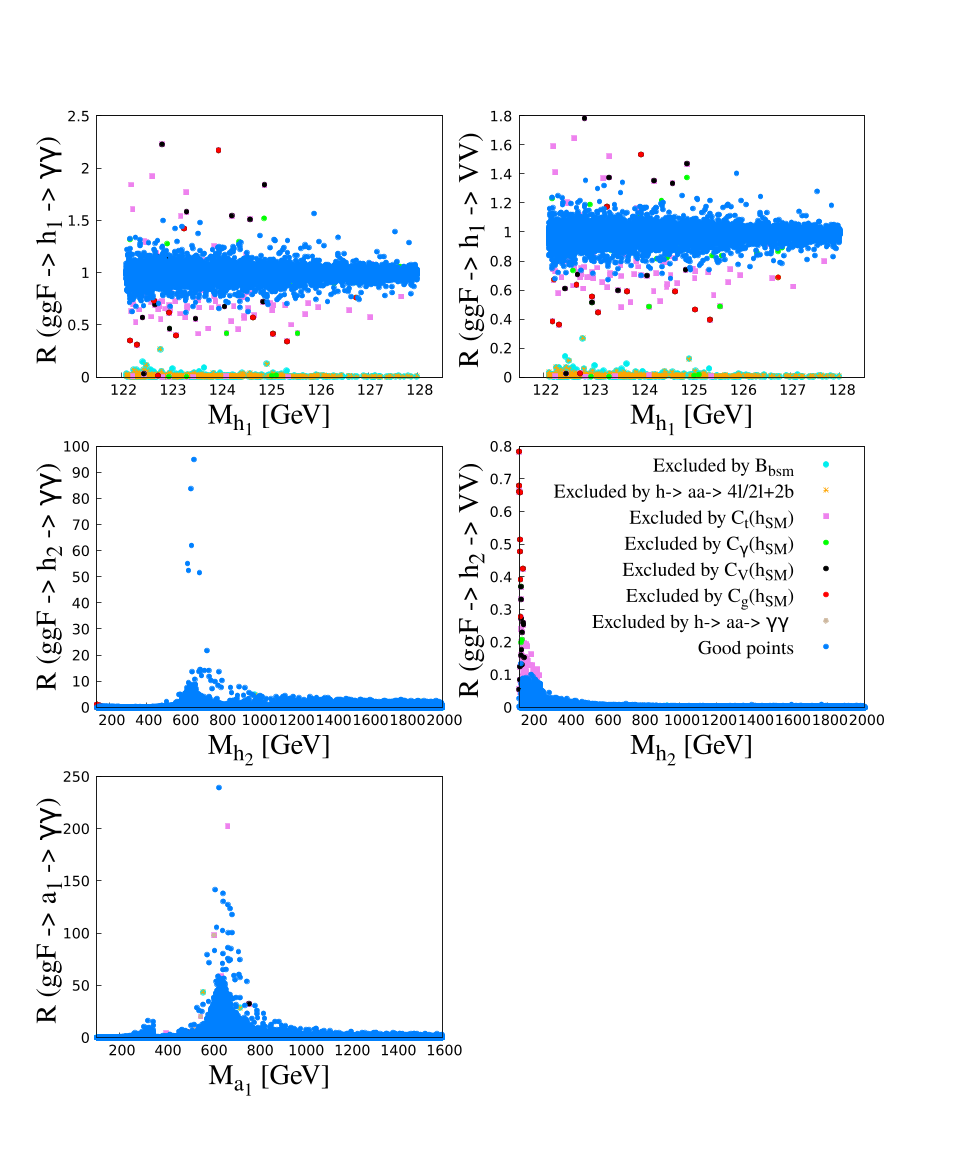}
    \caption{Reduced cross-section into $ \gamma \gamma$ and $ WW/ZZ$ versus the lightest CP-even Higgs mass $M_{h_{1}}$ (upper panels), the next to lightest CP-even Higgs mass $M_{h_{2}}$ (middle panels) and the lightest CP-odd Higgs mass $M_{a_{1}}$ (lower panels) with $\mathbf{A}_{0}\neq \mathbf{A}_{\protect\lambda
}\mathbf{\neq A}_{\protect\kappa }$. }
    \label{REDh1h2a1case2}  
\end{figure*}

Finally, the bottom row of Fig.~\ref{REDh1h2a1case2} presents the reduced cross-sections: $R(ggf \rightarrow a_1 \rightarrow \gamma \gamma)$ (left). The value fluctuates and reaches a maximum close to 250 at $m_{a_{1}} = 630$ GeV. The ruled-out points violate the limits on $C_{t}(h_{SM})$, $C_{\gamma}(h_{SM})$, $h \rightarrow aa \rightarrow 4l/2l+2b $, and $h \rightarrow aa \rightarrow \gamma \gamma $. 

\section{Discussion and Conclusion} \label{conc}
The results in Sec.~\ref{sc:res} show that the NMSSM with semi-constrained GUT boundary conditions can account for theoretical and experimental constraints. It is known that the NMSSM can admit either $h_1$ or $h_2$ (the lightest or second to lightest CP-even Higgs) as the SM-like Higgs. However, we observe that under the considered ranges for input parameters (mentioned in Sec.~\ref{sec:scans}) and the imposed constraints, the SM-like Higgs boson turned out to be the lightest CP-even Higgs ($h_1$). Indeed, expanding the range of parameters may well result in regions where $h_2$ is the SM-like. However, our choices for the scanned ranges were motivated by minimizing the fine-tuning in the model, which occurs in the large $\lambda$, and low $\tan{\beta}$ regime as explained in Sec.~\ref{sec:core1} and the relevant literature. Alas, the absence of points where $h_2$ is SM-like could be due to the limitations of the random scanning technique that was used in this study.

More importantly, we have systematically and thoroughly analyzed the Higgs sector of the model and how it is affected by recent constraints from the LHC on the SM-Higgs couplings. In particular, the predictions of the model related to: reduced couplings, doublet-singlet components, branching ratios to bosons, and reduced cross-sections. We limited our attention to the particles: $h_1, h_2$ and $a_1$, as they are more relevant to current efforts in exploring BSM physics at the LHC.
The results show that the Higgs sector has been impacted by aforementioned constraints.

We observe that the limits on the decays of the CP-odd Higgs $a_1$, and on the Higgs couplings to the top quark, and on the branching ratio of the SM-like Higgs to new physics have the strongest impact on the parameter space of the model.
As we map those constraints onto the parameter space, we find that regions where both $h_2$ and $a_1$ are less than 400 GeV are restricted by a combination of the previously mentioned limits. 

Most notably, for $h_2$, regions where its mass is below 300 GeV, its singlet component is below 0.85, its reduced couplings have the values: $|C_U| > 0.3$, or $C_g > 0.3$, or $C_{\gamma} > 0.5$, or $|C_V| > 0.3$, and its reduced cross-section to vector bosons is above 0.1, are ruled-out from the model due to the fact that such regions of parameter space are associated with violations of LHC constraints on the SM-like Higgs couplings to bosons and fermions. This suggests that $h_2$ should be mostly singlet for the model to be viable within the discussed regime. On the other hand, the results show that $a_1$ is also mostly singlet for regions where its mass is below 400 GeV. For such a range of mass, the most notable exclusion region is clearly seen in the results of its coupling to down-type quarks $|C_D| > 0.5$. There, most of the LHC constraints discussed are violated, especially the limits on $|C_{\gamma}(h_{SM})|$, and $B_{BSM}(h_{SM})$. Both $h_2$ and $a_1$ have large branching ratio to a pair of photons in the low-mass range.  

All in all, the NMSSM constitutes a good candidate for BSM physics, and we have provided an up-to-date analysis of its Higgs sector and its bosonic decays. As the LHC continues to probe the SM-like Higgs boson and provide more precise measurements and constraints on its properties, the parameter space of the model will be affected. Our results, and the provided plots, can be used as a reference to explore such new constraints. We have shown that the most affected regions by recent LHC SM-like Higgs coupling measurements are those regions associated with $m_{h_{2}} < 300$ GeV, and $m_{a_{1}} < 400$ GeV. Such realization suggests a future study whereby we analyze the interplay between Higgs coupling constraints and how might that affect the discovery potential of light scalars in the Higgs sector of the NMSSM. 

\section*{Acknowledgement}
The authors would like to thank the Deanship of Scientific Research in King Saud University for funding and supporting this research through the initiative of DSR Graduate Students Research Support (GSR). Maien Binjonaid would like to thank Ulrich Ellwanger for useful communications regarding NMSSMTools, and Roman Nevzorov for providing critical feedback on the manuscript of the paper.

\bibliographystyle{ws-mpla}
\bibliography{Higgs1}

\section*{Appendix}
Here we present a representative point that passes all experimental constraints specified earlier (apart from $(g-2)_{\mu}$), and where the lightest pseudoscalar is light and mostly singlet.

\begin{center}
\begin{tabular}{||c||c||cc}
\hline\hline
\textbf{Entry} & \textbf{value} & \textbf{Entry} & \multicolumn{1}{||c||}{%
\textbf{value}} \\ \hline\hline
$m_{0}(\text{GUT})\ $ & $2500$ $GeV$ & $h_2$ singlet component & \multicolumn{1}{||c||}{$0.628$} \\ 
\hline\hline
$m_{1/2}(\text{GUT})\ $ & $3100$ $GeV$ & $m_{h_{3}}$ & \multicolumn{1}{||c||}{$1957.2$ $%
GeV$} \\ \hline\hline
$A_{0}(\text{GUT})\ $ & $-90$ $GeV$ & $m_{a_{1}}$ & \multicolumn{1}{||c||}{$64.7$ $GeV$}
\\ \hline\hline
$A_{\lambda }(\text{GUT})$  & 0 & $%
a_1$ singlet component & \multicolumn{1}{||c||}{$0.9995$} \\ \hline\hline
$A_{\kappa }(\text{GUT})$  & $-100$ $GeV$ & $m_{a_{2}}$ & 
\multicolumn{1}{||c||}{$1937.7$ $GeV$} \\ \hline\hline
$\tan \beta(M_Z) $ & $1.6$ & $m_{h^{\pm }}$ & \multicolumn{1}{||c||}{$1936.4$ $GeV
$} \\ \hline\hline
$\lambda $ $\left( _{SUSY}\right) $ & $0.51$ & $Br\left( h_{2}\rightarrow
W^{+}W^{-}\right) $ & \multicolumn{1}{||c||}{$5.75\times 10^{-4}$} \\ 
\hline\hline
$\kappa $ $\left( _{SUSY}\right) $ & $0.41$ & $Br\left( h_{2}\rightarrow
ZZ\right) $ & \multicolumn{1}{||c||}{$2.87\times 10^{-4}$} \\ \hline\hline
$\mu _{eff}\ $\ $\left( _{SUSY}\right) $ & $1200$ $GeV$ & $Br\left(
h_{2}\rightarrow \gamma \gamma \right) $ & \multicolumn{1}{||c||}{$%
3.98\times 10^{-6}$} \\ \hline\hline
$m_{H_{u}}\ \left( _{GUT}\right) $ & $4.09\times 10^{7}$ $GeV$ & $Br\left(
h_{2}\rightarrow Z\gamma \right) $ & \multicolumn{1}{||c||}{$1.39\times
10^{-6}$} \\ \hline\hline
$m_{H_{d}}\ \left( _{GUT}\right) $ & $5.89\times 10^{5}$ $GeV$ & $Br\left(
h_{2}\rightarrow h_{1}h_{1}\right) $ & \multicolumn{1}{||c||}{$3\times
10^{-4}$} \\ \hline\hline
$m_{S}\ \left( _{GUT}\right) $ & $7.36\times 10^{6}$ $GeV$ & $Br\left(
h_{2}\rightarrow a_{1}a_{1}\right) $ & \multicolumn{1}{||c||}{$0.11$} \\ 
\hline\hline
$m_{\widetilde{t}_{1}}$ & $3227.6$ $GeV$ & $Br\left( h_{2}\rightarrow
Za_{1}\right) $ & \multicolumn{1}{||c||}{$5.88\times 10^{-2}$} \\ 
\hline\hline
$m_{\widetilde{g}}$ & $6387.6$ $GeV$ & $Br\left( a_{1}\rightarrow \gamma
\gamma \right) $ & \multicolumn{1}{||c||}{$2.63\times 10^{-4}$} \\ 
\hline\hline
$m_{\widetilde{\chi }_{1}^{0}}$ & $1201.8$ $GeV$ & $R\left( gg\rightarrow
h_{2}\rightarrow VV^{\ast }\right) $ & \multicolumn{1}{||c||}{$2.03\times
10^{-4}$} \\ \hline\hline
$m_{\widetilde{\chi }_{1}^{\pm }}$ & $1211.3$ $GeV$ & $R\left( gg\rightarrow
h_{2}\rightarrow \gamma \gamma \right) $ & \multicolumn{1}{||c||}{$1.12$} \\ 
\hline\hline
$m_{h_{1}}$ & $122.5$ $GeV$ & $R\left( gg\rightarrow a_{1}\rightarrow \gamma
\gamma \right) $ & \multicolumn{1}{||c||}{$3.83\times 10^{-4}$} \\ 
\hline\hline
$m_{h_{2}}$ & $1913.5$ $GeV$ & $\Omega h^{2}$ & 
\multicolumn{1}{||c||}{$0.1199$} \\ \hline\hline
\end{tabular}
\end{center}

\end{document}